\documentclass{aa}  

\usepackage[breaklinks=true,colorlinks,citecolor=blue]{hyperref}
\usepackage{graphicx,animate}
\usepackage{enumerate}
\usepackage{epsfig}
\usepackage{color}
\usepackage{txfonts}
\usepackage{natbib}
\usepackage{multirow}
\usepackage{ulem} 
\usepackage{amssymb}

\usepackage{movie15}

\usepackage{xcolor}

\definecolor{ao}{rgb}{0.0, 0.5, 0.0}

\newcommand{\kmps}{\rm km~s\ensuremath{^{-1} }\,}

\newcommand{\Msun}{M\ensuremath{_\odot}}

\newcommand{\Gaia}{{\it Gaia}\,}

\newcommand{\Fe}{\ensuremath{\rm Fe}\,}

\newcommand{\aFe}{\ensuremath{\rm [\alpha/Fe]}\,}

\newcommand{\dfde}{\ensuremath{\rm \nabla_E [Fe/H]}\,}
\newcommand{\dfdr}{\ensuremath{\rm \nabla_R [Fe/H]}\,}
\newcommand{\dfdrg}{\ensuremath{\rm \nabla_{R_g} [Fe/H]}\,}

\newcommand{\MnFe}{\ensuremath{\rm [Mn/Mg]}\,}
\newcommand{\AlFe}{\ensuremath{\rm [Al/Fe]}\,}
\newcommand{\MgFe}{\ensuremath{\rm [Mg/Fe]}\,}
\newcommand{\FeH}{\ensuremath{\rm [Fe/H]}\,}

\newcommand{\rmax}{\ensuremath{R_{max}}}
\newcommand{\rmin}{\ensuremath{R_{min}}}
\newcommand{\zmax}{\ensuremath{Z_{max}}}
\newcommand{\ecc}{\ensuremath{\varepsilon}}
\newcommand{\cir}{\ensuremath{\lambda}}

\begin{document} 

\title{Dead man tells tales: metallicity distribution of the Milky Way stellar halo reveals the past of the GSE progenitor galaxy}
 
\titlerunning{GSE-progenitor}

\author{Sergey Khoperskov$^{1}$, Ivan Minchev$^{1}$, Matthias Steinmetz$^{1}$, Julien Marabotto$^{2,1}$,  \\ Georges Kordopatis$^{3}$, Jeicot Delgado Gomez$^{4,1}$, Noam Libeskind$^{1}$
}

\authorrunning{Khoperskov et al}

\institute{ $^{1}$ Leibniz Institut f\"{u}r Astrophysik Potsdam (AIP), An der Sternwarte 16, D-14482, Potsdam, Germany\\
$^{2}$ Department of Physics, University of Surrey, Guildford, Surrey, GU2 7XH, UK \\
$^{3}$ Université Côte d'Azur, Observatoire de la Côte d'Azur, CNRS, Laboratoire Lagrange, 06000, Nice, France \\
$^{4}$ Institut für Physik und Astronomie, Universität Potsdam, Karl-Liebknecht-Str. 24/25, 14476, Potsdam, Germany
}

\abstract{The Gaia-Sausage-Enceladus~(GSE) stands out as the largest known ancient accretion event in the Milky Way~(MW) history. Despite this significance, the parameters of its progenitor galaxy are still poorly constrained. We identify GSE stars from the APOGEE DR17 using Gaussian mixture models and recover a negative radial metallicity gradient for the GSE debris within the MW stellar halo, with a magnitude of $\approx -0.014^{-0.002}_{-0.022}$ dex/kpc.
We argue that this gradient reflects the radial metallicity gradient of the GSE galaxy progenitor before it was disrupted by the MW. By investigating the cosmological HESTIA simulations and $N$-body models of galaxy mergers, we constrain the radial metallicity gradient of the GSE-progenitor to be $\approx -0.1^{-0.06}_{-0.15}$ dex/kpc. We, therefore, propose that a chemical tagging of accreted stars using their integrals of motion, although they are not conserved during mergers, provide essential information about the structure and the past of systems accreted onto the MW.
}


\maketitle

\section{Introduction}\label{sec::intro}

According to the $\Lambda$ cold dark matter cosmology, galaxies grow via hierarchical mergers of smaller systems~\citep{1985Natur.317..595F, 1991ApJ...379...52W, 1993MNRAS.262..627L, 1993MNRAS.264..201K, 2005Natur.435..629S}. However, this appears not to have been the case for the Milky Way~(MW) galaxy since at least $90\%$ of its stellar mass appears to be formed in situ. Recent estimates constrain the mass of the MW stellar disk in the range of $6-8\times 10^{10}$~\Msun~\citep{2013ApJ...779..115B, 2014ApJ...794...59K, 2017MNRAS.465...76M, 2019ApJ...871..120E}. At the same time, several methods provide estimates of the stellar halo mass from $4-7\times 10^8$~\Msun~\citep{2016ARA&A..54..529B,2017MNRAS.465...76M} to $1.4 \pm 0.4 \times 10^9$~\Msun~\citep{2019MNRAS.490.3426D}. These measurements include  the major contributions from the ongoing merger with the Sgr dwarf~\citep{1994Natur.370..194I, 1995ApJ...451..598J, 2004ApJ...601..242M}, stars acquired by the MW during past merger events~\citep{1999MNRAS.307..495H, 1999Natur.402...53H, 2016ApJ...821....5D} and in-situ stellar populations heated by mergers~\citep{2007Natur.450.1020C, 2010ApJ...714..663D, 2017ApJ...845..101B,2019A&A...632A...4D, 2020MNRAS.494.3880B}. Therefore, as expected from a large number of galaxy formation simulations~\citep{2006MNRAS.365..747A, 2005ApJ...635..931B,2010MNRAS.406..744C, 2015MNRAS.454.3185C, 2011MNRAS.416.2802F, 2008ApJ...689..936J, 2010MNRAS.404.1711P, 2009ApJ...702.1058Z, 2020MNRAS.497.1603G, 2021MNRAS.503.5846R, 2022arXiv220604522K, 2022arXiv220604521K}, the MW stellar halo is made of heated disc stars and stars from smaller stellar systems disrupted by the tidal forces of the MW.

Among several already accreted systems~(excluding the Sgr and LMC/SMC) discussed so far in the literature, the most significant is the Gaia-Sausage-Encelladus~(GSE), which tip was known since works by \cite{2000AJ....119.2843C, 2010A&A...511L..10N, 2011A&A...530A..15N, 2012ApJ...746...34B}. However, its presence becomes evident thanks to the \Gaia data \citep{2018MNRAS.478..611B,2018ApJ...863..113H,2018Natur.563...85H}. What is known about a dwarf galaxy so far whose merger remnant is seen as GSE stars in the MW halo? 

\cite{2020A&A...642L..18K} suggest that the GSE-progenitor was a disc dwarf galaxy that fell in on a retrograde orbit~\citep[see also][]{2021ApJ...923...92N}. Their $N$-body model assumes a relatively massive dwarf galaxy~($>10^{10}~\Msun$), implying the importance of dynamical friction on its stellar debris structure. There are several more comprehensive studies allowing us to estimate the mass of the GSE-progenitor more precisely. For instance, \cite{2019MNRAS.482.3426M} and \cite{2020MNRAS.492.3631M} analysed dynamics and elemental abundances of stars from the APOGEE data, resulting in a stellar mass estimate of $3-10\times 10^8\Msun$. Alternatively, mass estimates utilizing dynamical modelling~\citep{2021ApJ...923...92N}, mass-metallicity relations ~\citep{2020MNRAS.497..109F, 2020MNRAS.498.2472K, 2020ApJ...901...48N,2021MNRAS.500.1385H} or chemical evolution modeling \citep{2018Natur.563...85H, 2019MNRAS.487L..47V} suggest the range from $4 \times 10^8$ to $7\times 10^9\Msun$ for the stellar mass of the GSE-progenitor. In several studies, the virial mass of the GSE-progenitor was constrained as $\sim 10^{11}\Msun$~ \citep{2018MNRAS.478..611B, 2020MNRAS.498.2472K,2021ApJ...923...92N, 2023arXiv230600770C}.

Although the GSE progenitor galaxy has been accreted already about $9-11$ ago~\citep{2018Natur.563...85H, 2019MNRAS.482.3426M, 2019NatAs...3..932G,2021NatAs...5..640M}, there was enough time to evolve chemically to reach metallicity of about $\approx -0.5$~dex with the mean value of $\approx -1.2$~dex. The models of its chemical evolution suggest that the star formation inside this dwarf galaxy should be more effective compared to the present-day satellites of the MW because of a higher \aFe ratio in the same metallicity range~\citep{2021ApJ...923..172H}. There is also evidence of a peak of star formation just before it was accreted~\citep{2019NatAs...3..932G,2021ApJ...923..172H}, as it is predicted in simulations of galaxy mergers~\citep{2008A&A...492...31D, 2014MNRAS.442L..33R, 2021MNRAS.506..531D, 2022MNRAS.516.4922R}.

Therefore, the GSE progenitor was a dwarf galaxy with a stellar mass formed on a scale of $\approx 2-5$~Gyr before its accretion onto the MW. However, if that had not happened, since the GSE progenitor had a substantial amount of gas~\citep{2020MNRAS.497.1603G,2023MNRAS.tmpL..32C}, at $z=0$ it would probably be more massive than the present-day LMC or M32 satellite galaxies. However, since its evolution stopped at the time of accretion, the GSE progenitor is not expected to be identical to the dwarf galaxies that survived to the present day. Nonetheless, the Local Group~(LG) dwarfs are the obvious candidates for comparison, especially if they shut down their star formation early on. It is believed that dwarf galaxies have a complex structure of their stellar populations where more metal-rich stars tend to be more compact compared to more radially extended metal-poor stars~\citep{2001AJ....122.3092H, 2004ApJ...617L.119T, 2006A&A...459..423B, 2011ApJ...727...78K, 2011MNRAS.411.1013B, 2012MNRAS.422...89M, 2012ApJ...761L..31B}, which can be explained if their inner parts start to form earlier, thus evolving longer and reaching higher metallicites, while the star formation in the outer parts is either start later or less efficient in enriching surrounding ISM~\citep{2006ApJ...641..785K,2008MNRAS.389.1111V,2013MNRAS.434..888S,2016MNRAS.456.1185B,2016MNRAS.457.1299K,2018A&A...616A..96R,2009MNRAS.395.1455S,2021MNRAS.501.5121M}. In this scenario negative radial metallicity gradients are a common feature of dwarf galaxies. We stress, however, that literature suggests that observed metallicity gradients can be the result of the dwarf galaxies' evolution under the influence of more massive host galaxies affecting their internal structure, namely through gas stripping, tidal fields and SF activity via contraction of their gas~\citep[see, e.g.][]{2016MNRAS.457.1931S,2016ApJ...827L..23W,2021ApJ...909..139A,2021MNRAS.506..531D}. Nevertheless, most of the LG dwarf galaxies do reveal (usually negative) radial metallicity gradients~\citep{2011ApJ...727...78K,2013ApJ...767..131L,2022A&A...665A..92T}. 

According to a recent compilation by \cite{2022A&A...665A..92T}, the LG dwarf galaxies with the oldest stellar populations, such as Sextans, Pegasus, Tucana, Draco, and others, exhibit radial metallicity gradients ranging from approximately $-0.5$ to $-0.1$~dex/kpc. This suggests that strong metallicity gradients in dwarf galaxies can be established very early in their formation process~\citep{2018A&A...616A..96R}. Given this understanding, it is reasonable to assume that the wide range of metallicities observed among the GSE stars signifies the presence of spatial abundance variations within its galaxy progenitor, including a potential radial metallicity gradient. However, recovering this gradient from the present-day, phase-mixed MW stellar halo is challenging due to the complexities introduced by the low number of its stars, completeness issue and selection function features of different MW halo surveys.

Recently, using HESTIA cosmological simulations of the LG, \cite{2022arXiv220604522K} showed that the radial structure of accreted systems can be traced back as a function of the integrals of motion of stellar merger debris. In particular, it was demonstrated that inner, more metal-rich parts of dwarf galaxies tend to sink towards the innermost parts of M31/MW analogues and have lower total energy, while outer more metal-poor regions of disrupted dwarf galaxies mainly appear to be less bound to the host galaxy with higher energies~\citep[see also][]{2021ApJ...923...92N,2022ApJ...937...12A,2022arXiv220605491K,2023A&A...673A..86P,2022MNRAS.517L.138O,2023ApJ...944..169D}. Therefore, the radial abundance and stellar age gradients of progenitors of accreted galaxies could be recovered from the corresponding gradients as a function of the integrals of motion in the MW stellar halo.

However, \cite{2023ApJ...944..169D} showed that the abundance variations in both APOGEE and Galah datasets are not consistent with this scenario, which can be explained if the GSE populations are made of several accreted systems. Indeed, following the identification of the GSE merger debris in the \Gaia data, a diverse range of chemo-kinematic substructures have been identified, claiming that each of them can be attributed to a unique accreted system~\citep[we refer to a compilation of most of the substructures in ][]{2022ApJ...926..107M}. The biggest challenge in the identification of different substructures is the demarcation lines or boundaries between them and the known populations~(GSE and in situ). The best illustration of this situation is given in \cite{2023MNRAS.520.5671H}~(see their Fig.~4), where straight lines~(compilation of selections from literature) demonstrate the boundaries between different groups. However, such behaviour of stellar remnants of accreted systems is not really predicted by simulations~\citep{2017A&A...604A.106J, 2019MNRAS.487L..72G, 2019MNRAS.484.4471F, 2022arXiv220604522K, 2022ApJ...934..172C, 2023A&A...673A..86P}, demonstrating a substantial overlap between different accreted debris and in situ stars in all possible combinations of chemo-kinematic spaces.

\begin{figure*}
\begin{center}
\includegraphics[width=0.99\hsize]{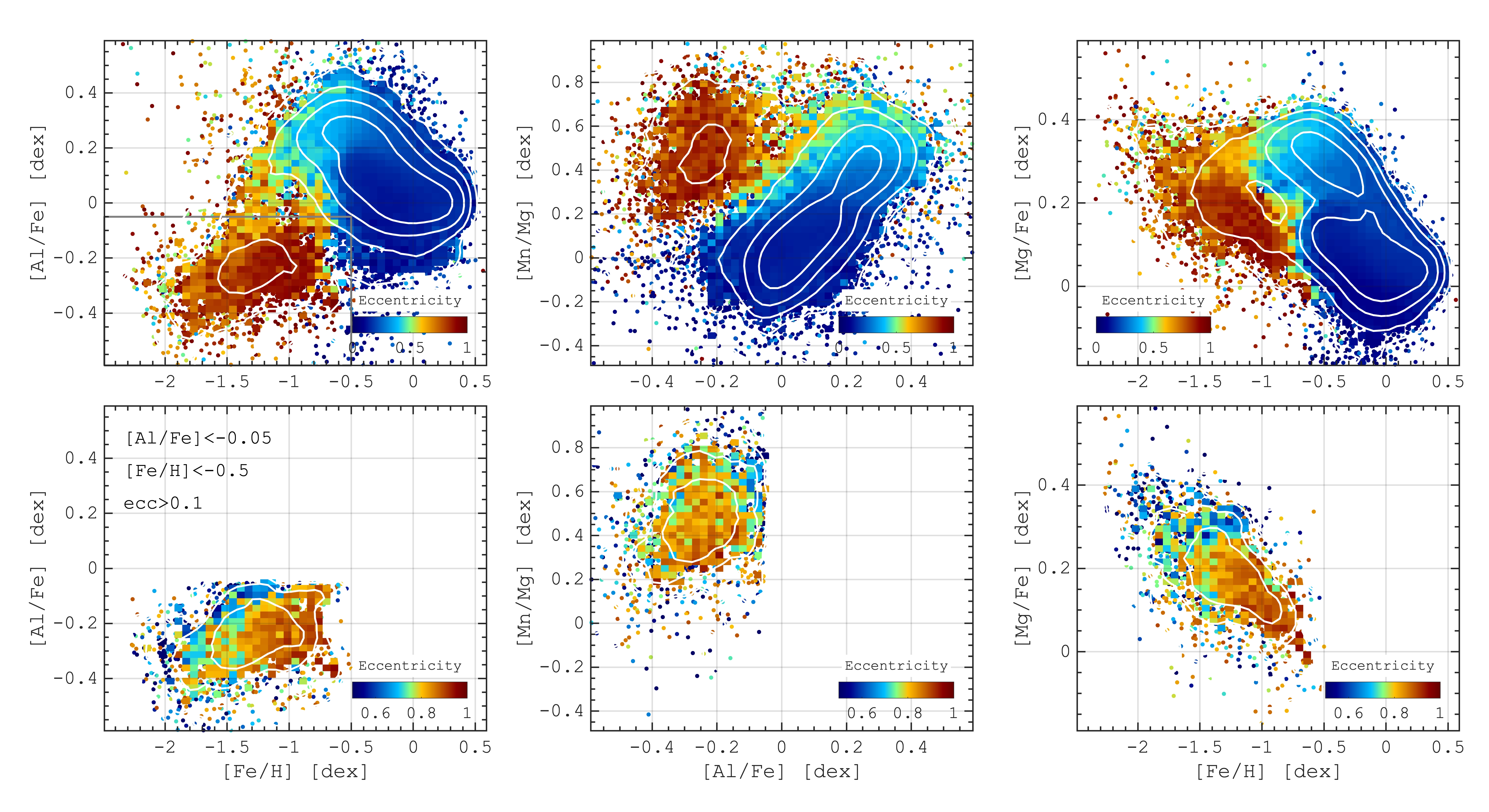}
\caption{Chemical abundance composition of the APOGEE sample. The white lines show the density contours in a $\log$ scale while the background maps depict the mean orbital eccentricity. The top panels correspond to the entire input sample. The bottom panels show only stars selected $\FeH<-0.5$ and $\AlFe<-0.05$ and $\ecc>0.1$ which are predominantly populated by the high-eccentricity accreted stars with some contamination of in situ stellar populations with lower-eccentricities. To highlight the contamination of the low-eccentricity stars, the colour range of the mean eccentricity is different in the bottom row relative to the top. The lower-eccentricity regions in the bottom are linked to the chemically defined thick disk~(high-$\alpha$ sequence), suggesting their genetic connection. }\label{fig::apogee_chem_select}
\end{center}
\end{figure*}

The lack of an objective approach in the identification of the MW stellar halo substructures has been revised by \cite{2022ApJ...938...21M}, who used non-supervised Gaussian Mixture models~(GMMs) in order to decompose the local stellar halo in the chemical abundance space. Among four prominent groups recovered by their models, only one can be classified as accreted -- GSE, while the remaining three are different known sub-populations of the in situ MW stars~(Aurora, 'spin-up'~\citep{2022MNRAS.514..689B} and Splash/Plume~\citep{2020MNRAS.494.3880B,2019A&A...632A...4D}). Therefore, so far, only one ancient and massive accreted system has been identified unambiguously. This is supported by the fact that the \Gaia colour-magnitude diagram shows a prominent bimodality~\citep{2018A&A...616A..10G,2018ApJ...863..113H} with only two clear sequences corresponding to the GSE and in situ formed halo components and without any strong signatures of additional components. From the simulation point of view, since the GSE is believed to be the most recent significant merger, it should be the most radially extended accreted structure inside the MW halo, with a substantial contribution in the extended solar vicinity, dominating over the less massive~(even earlier) mergers remnants. These should occupy the innermost parts of the MW~\citep{,2022arXiv220604522K} which are poorly covered by the present-day spectroscopic surveys. Finally, the stellar density break and the metallicity profiles in the MW halo are also in favour of a single major accretion event in the MW history~\citep{2013ApJ...763..113D, 2018ApJ...862L...1D}.

Negative radial metallicity gradients is a natural outcome of the evolution of dwarf galaxies, which, as theory suggests, can be traced in the stellar merger remnants of accreted systems. In this work, aiming to put constraints on parameters of the GSE galaxy progenitor, we explore the origin of the metallicity variations of the GSE merger debris and compare it to constrained $N$-body merger simulations and cosmological HESTIA simulations. The paper has the following structure. In Section~\ref{sec::data} we describe the observational data, in particular, the APOGEE data set and present our approach for the selection of the GSE members. In Section~\ref{sec::models} we present a new set of idealised $N$-body merger simulations and the HESTIA simulations. In Section~\ref{sec::results} we present the reconstruction of the GSE progenitor metallicity distribution. In Sections ~\ref{sec::discussion} and ~\ref{sec::summary},  we discuss and summarise our main findings, respectively.

\section{Selection of GSE stars}\label{sec::data}

\begin{figure*}
\begin{center}
\includegraphics[width=1.\hsize]{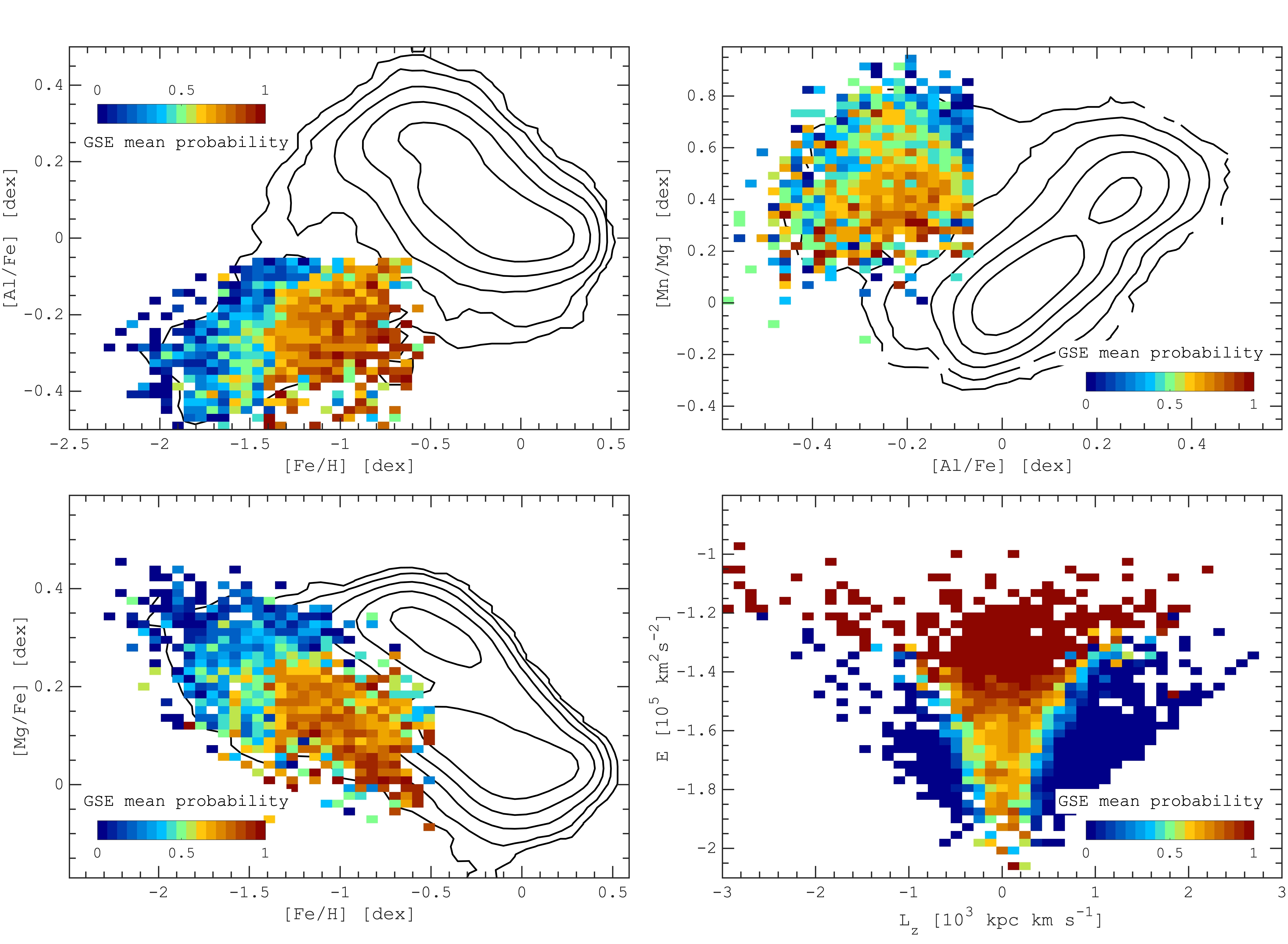}
\caption{Results of the GMM in the classification of the low-\FeH APOGEE stars onto accreted and in situ populations. The maps show the mean probability of stars being associated with accreted stars~(reddish colours). For the accreted stellar populations, the GMMs deliver a selection suggested for the GSE merger debris with little net rotation and, on average, lower \AlFe, \MgFe and \MnFe values. The in situ stars~(bluish colours) represent a low-\Fe tail of the early MW connected to the high-$\alpha$ sequence. The black contours are given for reference and depict the density distribution of the entire APOGEE dataset.}\label{fig::validation}
\end{center}
\end{figure*}

\subsection{Initial sample from APOGEE and \Gaia}

In this study, we used data from the \Gaia Data Release 3~\citep{2023A&A...674A...1G} and APOGEE DR17~\citep{2017AJ....154...94M,2022ApJS..259...35A}. The distances and their uncertainties were taken from the AstroNN value-added catalogue~\citep{2019MNRAS.483.3255L}, where we selected stars with $<20\%$ distance uncertainty. Similarly to \cite{2021MNRAS.500.1385H}, we chose to restrict our analysis to stars with surface gravity estimates of $\rm logg < 3.6$ and $\rm 3500 < T_{eff} < 5500$~K. For the radial velocity and proper motions uncertainties, we adopted quality cuts of $10$~\kmps and $20\%$, respectively. We also removed stars with any $\rm STARFLAG$ or $\rm ASPCAPFLAG$ flags. Moreover, we removed stars in fields targeted for known globular clusters or dwarf galaxies. We also required uncertainty of $< 0.2$ dex in the estimates of the following abundances: \MgFe, \AlFe, \FeH, and \MnFe. Since we are interested in understanding the properties of the local stellar halo sample, we restricted our analysis to stars located not farther than $5$~kpc from the Sun. This spatial selection also ensures small contamination from the bulge/bar region and possibly unknown accreted systems in the very centre of the MW.

The orbital parameters~(apocentres \rmax, pericentres \rmin, maximum vertical excursion relative to the MW mid-plane \zmax, eccentricity \ecc\, and circularity \cir), the angular momentum~($L_z$) and orbital energies~($E$) were calculated with the AGAMA software~\citep{2019MNRAS.482.1525V}, using the MW potential from \cite{2017MNRAS.465...76M}.

Our input dataset is shown in Fig.~\ref{fig::apogee_chem_select}~(top), where we present three chemical abundance planes colour-coded by the mean orbital eccentricity. In this figure, we can see three prominent groups: (1) an $\alpha$-rich, medium eccentricity, disc population, (2) an $\alpha$-poor, low eccentricity, disc population, and (3) a low-\FeH, high eccentricity, accreted population~\citep[][]{2015MNRAS.453..758H, 2020MNRAS.493.5195D, 2021MNRAS.508.1489F, 2021MNRAS.500.1385H, 2022MNRAS.510.2407B, 2023arXiv230300016F}. To select only accreted stars in the APOGEE dataset, first, we applied chemical abundance cuts to eliminate the vast majority of in-situ stars by removing those with $\AlFe>-0.05$ and $\FeH>-0.5$~(see also \cite{2022MNRAS.514..689B}). After this, we end up with stars belonging to the GSE, but also disc stars at high-\AlFe and high-\FeH. To remove the bulk of the thin disc stars we adopted the eccentricity cut of $0.1$. The remaining sample of 3632 stars is presented in the bottom row of Fig.~\ref{fig::apogee_chem_select}, which is also colour-coded with the mean orbital eccentricity; note, however, the change in the colour range to $0.5-1$. This presentation shows a prominent variation of the orbital motions of stars across chemical abundance planes: the upper part of the \AlFe-\FeH relation~(left) has systematically lower eccentricity, and the same effect is seen on the top of the \MnFe-\AlFe relation~(centre) and for \MgFe>0.2 in \MgFe-\FeH plot~(right). 

The presence of in-situ, in particular, inner disc, stars in the $\AlFe<0$ and $\FeH<-0.5$ region has been demonstrated by \cite{2021ApJ...923..172H} and \cite{2022MNRAS.514..689B}. This rises much more rapidly along \FeH and resides mostly above the accreted sequence across the entire range of \FeH. Similarly, for the \MnFe-\FeH plane, \cite{2021MNRAS.500.1385H} presented a chemical evolution model, where both accreted and in-situ stellar sequences start at $\AlFe<-0.5$ but the in-situ track always lies above the accreted one~(see their Fig.~2). Finally, a contamination of high-$\alpha$ in situ stars at \FeH<-1 is expected from the basic principles of galactic chemical evolution, where the early MW disc evolution should have started in the extremely low-metallicity ISM and has nearly constant~(or even increasing) $\alpha$-abundances~\citep{1995ApJS...96..175B, 1999MNRAS.309..533B, 2002AJ....124..931B, 2020A&A...636A.115D}. From Fig.~\ref{fig::apogee_chem_select}~(bottom) we conclude that the chemical abundance region, usually considered as accreted, is, in fact, contaminated by the metal-poor tail of the high-$\alpha$ disc population, which is highlighted by a slightly lower mean orbital eccentricity. 

\begin{figure*}
\includegraphics[width=1.0\hsize]{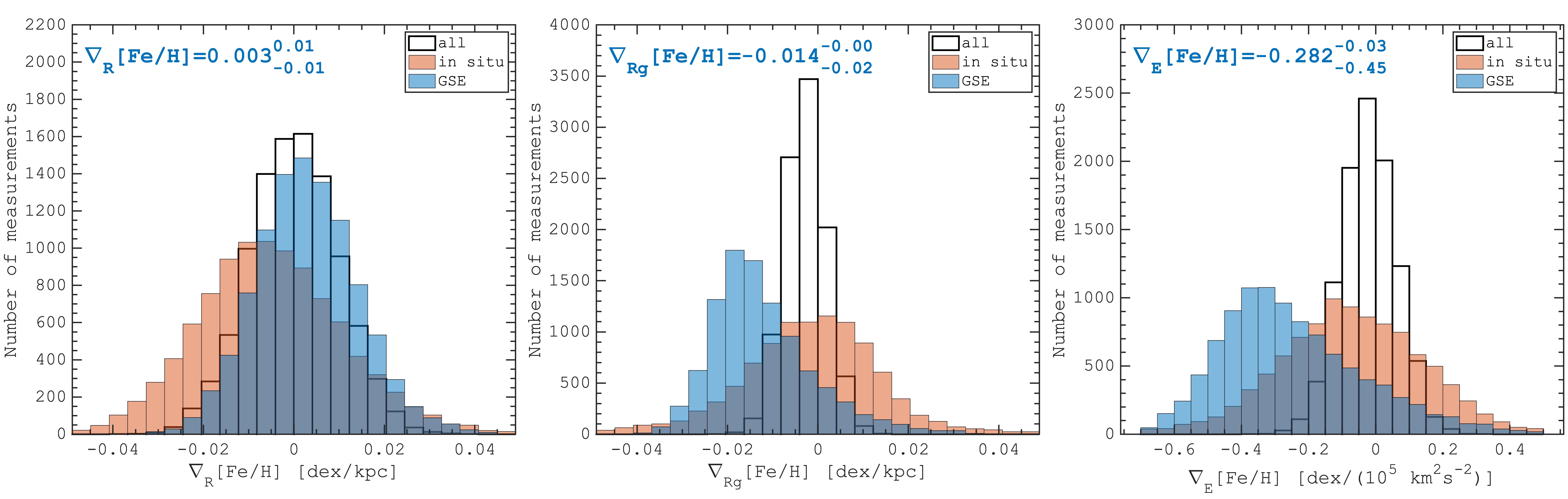}
    \caption{Probability distributions of the metallicity gradients of the GSE merger debris. The metallicity gradients were measured against instantaneous galactocentric distance ($\dfdr$, left), guiding radius ($\dfdrg$, centre) and total energy ($\dfde$, right) and are based on 10,000 GMM runs using the APOGEE dataset. Each measurement is done on a random sub-sample of 500 stars, classified by the GMM into GSE and in-situ components~(see more details in Sec.~\ref{sec::data}). The GSE merger debris shows a negative metallicity gradient as a function of the mean galactocentric distance and the total energy of stars inside the MW halo. In each panel, the median values of the metallicity gradients of the GSE debris with its upper and lower boundaries are displayed in blue. The presence of the negative metallicity gradients of the GSE debris suggests the negative metallicity gradient inside its galaxy progenitor before the merger.}
    \label{fig::gradients}
\end{figure*}

\begin{figure*}
\begin{center}
\includegraphics[width=1.0\hsize]{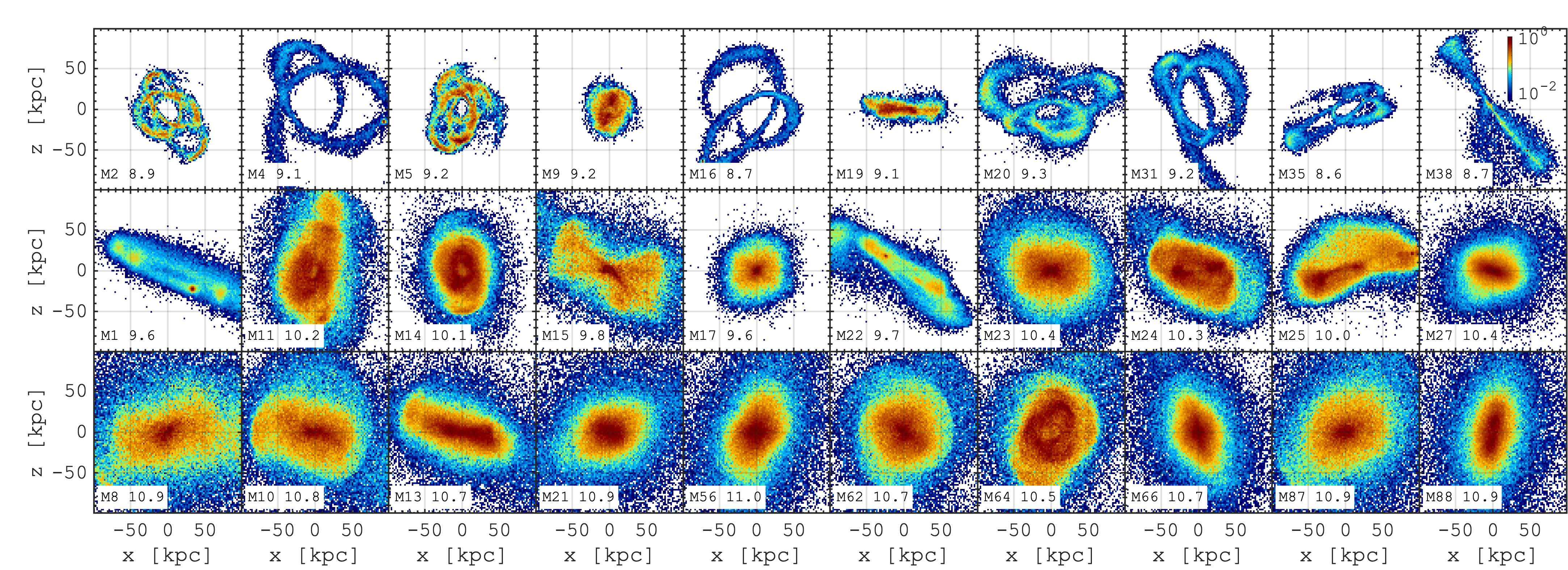}
\includegraphics[width=1.0\hsize]{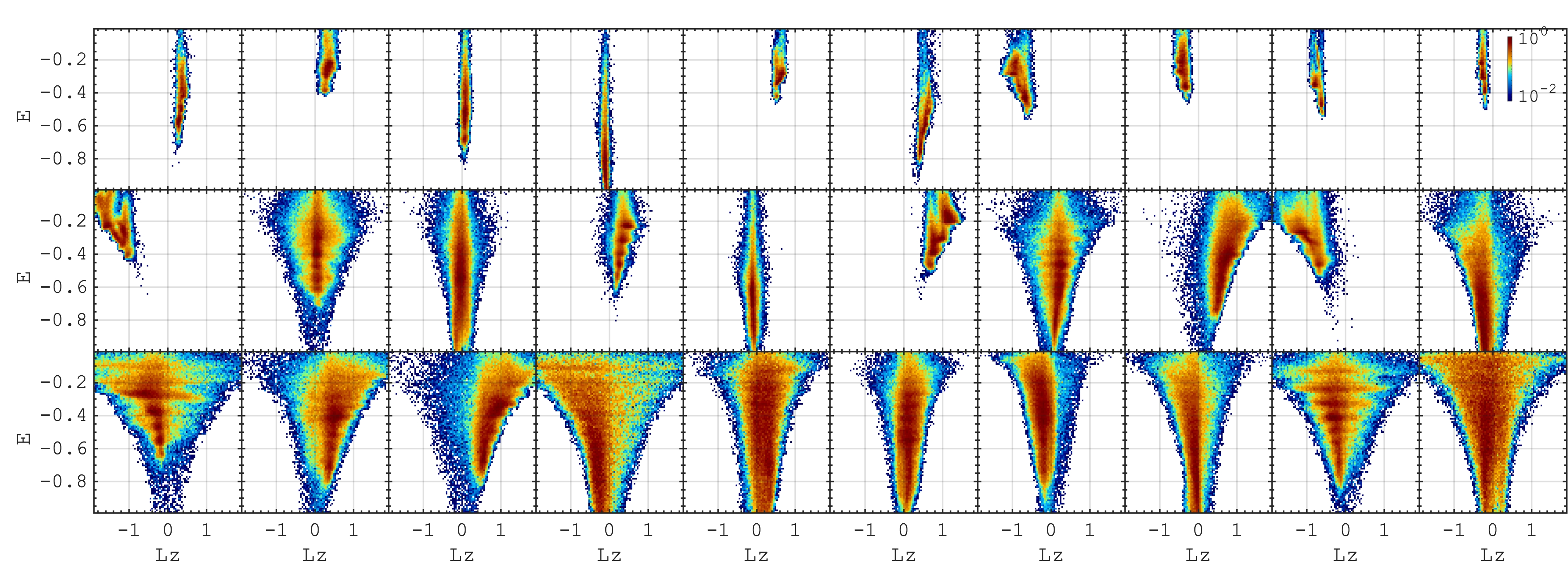}
\caption{Merger debris structure from $N$-body simulations considering $10$ Gyr evolution inside a self-consistent model of a MW-like disc galaxy. Each merger simulation considers the evolution of a single dwarf galaxy. {\it Top block of three rows:} Morphology of ten~(out of 1000) randomly-selected merger simulations, categorised according to the total mass~(see more details in Sec.~\ref{sec::models}). The top panel illustrates mergers within the mass range of $10^{8.5}-10^{9.5}\Msun$~(top),  $10^{9.5}-10^{10.5}\Msun$~(middle) and  $10^{10.5}-10^{11.5}\Msun$~(bottom). The model id and corresponded mass of accreted galaxy~(in units of $\rm \log_{10}(M/\Msun)$), are marked in each panel. {\it Bottom block of three rows:} Energy-angular momentum space for the same galaxy merger simulations. The maps highlight diverse morphological structures and energy-angular momentum distributions resulting from the mergers as a function of their total mass. Mergers with a total mass larger than $>10^{10}~\Msun$ result in debris similar to the GSE. In each paned the density distribution is normalized by maximum value.}\label{fig::MMs_morphology}
\end{center}
\end{figure*}

\begin{figure*}
\begin{center}
\includegraphics[width=1\hsize]{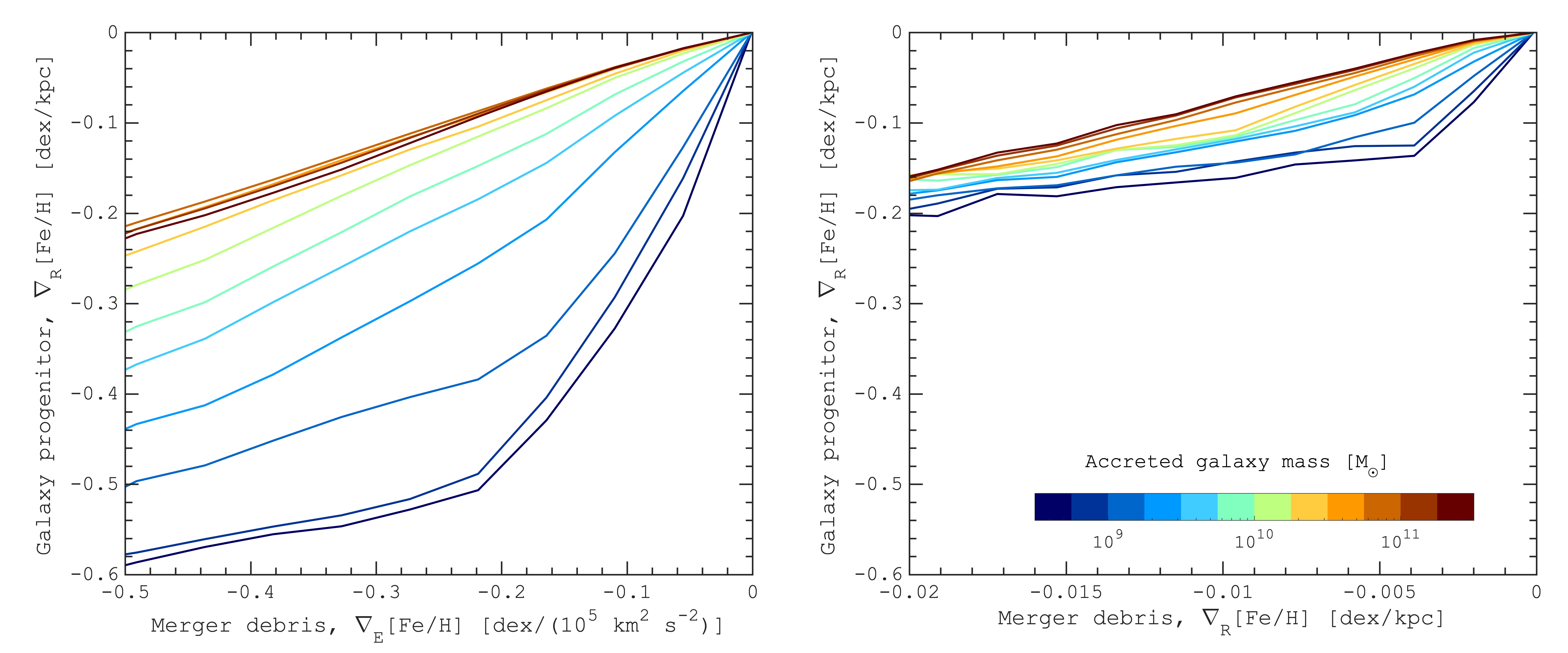}
\caption{Relation between the radial metallicity gradient of a stellar merger debris~(\dfde in left and \dfdr in right) and the radial metallicity gradient of its galaxy progenitor~(\dfdr) in 1000 $N$-body merger simulations, where for each simulation $30$ random  radial metallicity gradients~(from $-1$~dex/kpc to $0$~dex/kpc) were adopted. The lines represent the trends averaged in bins of the total mass of the galaxy progenitor, as it is marked by the colour bar in the right panel. A well-defined positive correlation is seen, where stronger metallicity gradients in the progenitor lead to correspondingly stronger gradients in merger debris.}
\label{fig::MMs_grads}
\end{center}
\end{figure*}

\subsection{Gaussian mixture models: cleaning accreted sample from the in-situ contamination}

As discussed in the introduction, sharp definitions of accreted and in-situ stars are not supported by simulations and can lead to significant biases in the resulting sample. To address this issue, we adopted an approach similar to \cite{2022ApJ...938...21M}, wherein we utilised GMMs to identify accreted stars in our sample~(see bottom panel in Fig.~\ref{fig::apogee_chem_select}). This data-driven approach allows for a more robust and flexible classification of stars, avoiding the imposition of sharp boundaries and taking into account the inherent complexity of stellar populations.

We adopted the following procedure: from a reduced sample of 3632 stars, we randomly selected 500 stars (we also experimented with samples of 1000 or 2000, yielding similar results). We employed an unsupervised GMM with only two components, namely GSE and in situ, and repeated this process 10,000 times. To address potential degeneracy due to correlations between the abundance trends and kinematic characteristics of stellar populations, we used a relatively small set of input parameters: \FeH, eccentricity~(\ecc), and \zmax. Each GMM-based classification assigns a probability (ranging from 0 to 1), indicating the likelihood of a star belonging to one of the two components.

As GMM is an unsupervised method, we do not control the fractions of the two components or the positions of mean values along any coordinates. Additionally, the GMMs cannot guarantee that the same component (accreted or in-situ) consistently corresponds to the same group during classification. To assign the classification probabilities to the accreted and in-situ components, we rely on the assumption that a group with a higher median angular momentum $\rm |L_z|$ is in situ, while a group with a lower median $\rm |L_z|$ corresponds to the accreted component.

As a result of this procedure, we obtained 110-170 probability values for a given star, representing the likelihood that this star corresponds to the GSE or the in-situ component. The distributions of the median values of these probabilities are depicted in Fig.~\ref{fig::validation}, where reddish colours (close to $1$) indicate a high contribution of the GSE stars, and bluish colours (close to $0$) highlight regions populated by the in-situ stars. The figure clearly demonstrates that our GMM-based selection of in-situ and accreted GSE stars, yields reasonable behaviour for both components in the chemical abundance planes, as we discussed in the previous section.

Additionally, in the bottom right panel of Fig.~\ref{fig::validation}, we present the $\rm E-L_z$ plane, where GSE stars dominate at high energies but also extend down to the lowest values of energy, consistent with simulations~\citep{2017A&A...604A.106J,2019MNRAS.487L..72G,2022arXiv220604522K}. Conversely, according to our definition, the in-situ stars prevail in a region with higher angular momentum but also exhibit a prominent retrograde motion, similar to the behaviour of spin-up stars within the metallicity range from $-1.5$ to $-1$~\citep{2022MNRAS.514..689B}. In our case, such behaviour should be even more pronounced, as the bulk of the in-situ stars have even lower metallicities. Our selection of the in-situ stars down to $\FeH\approx -2.2$ also aligns well with recent findings by \cite{2022ApJ...941...45R}, who used low-resolution XP spectra from \Gaia DR3 to identify non-rotating stellar populations with metallicities below $\FeH = -2$. We recall that our approach does not rely on sharp boundaries between different populations, allowing for potential overlap between the GSE and in-situ stars in all presented chemo-kinematic spaces.

\subsection{Metallicity gradient(s) of the GSE debris}

Since we ran the GMM 10,000 times on randomly extracted samples of stars, we could now calculate the metallicity gradient for each of these sub-samples. This extensive process allows us to examine the metallicity gradient variations across various subsets of stars where the probabilities delivered in each run are used as weights~(or inverse probability for the in-situ stars) in the regression analysis. Therefore, for the metallicity gradients calculation, we do not limit our sample of the GSE~(or in situ) stars to any probability value. 

In Fig.~\ref{fig::gradients} (left), we examine the distribution of radial metallicity gradients (\dfdr, where the index $R$ denotes the galactocentric distance). We also calculate the gradients for all stars in our sample, represented by black lines, but we analyse the radial metallicity gradients for the GSE stars (blue) and in-situ stars (red) separately. We observe that none of these groups display a significant gradient as a function of galactocentric distance. This is evident from the fact that the median values of the gradient distributions are centred around the zero point. This implies that, on average, there is no systematic variation in metallicity with respect to the distance from the galactic centre, for either the entire star sample or the GSE and in-situ subgroups. Although our sample is very local~(within $5$ kpc from the Sun), this conclusion is in agreement with some recent results suggesting the lack of halo metallicity gradient on a larger scale~\citep{2015ApJ...809..144X,2016MNRAS.460.1725D,2019ApJ...887..237C}. 

The MW halo stars, including those in our sample, are known for their high eccentricities, indicating that they predominantly move on radial orbits within the Galaxy. These stars undergo rapid oscillations between their apocentres and pericentres, where they spend relatively more time compared to other regions of their orbits. Hence, by using the mean galactocentric radius instead of its instantaneous value, we can gain a better understanding of the overall metallicity trends along their orbits even in a small and relatively local sample of stars. 

In the following, we consider that the mean (or guiding) galactocentric radius, calculated as $R_g = 0.5 \times (\rmax + \rmin)$, is a valuable approach to analyse the metallicity gradients for halo stars. Although this estimation of the mean value between apocentre and pericentre is relatively simple, it provides insights into the average metallicity at a more representative radial location. More sophisticated orbit-averaged calculations might not necessarily yield more realistic values for metallicity gradients, as we do not account for the radial variations in the halo shape and its evolutionary changes over time~\citep{2015ApJ...802..128G,2021ApJ...919..109G,2021Natur.592..534C,2021MNRAS.501.2279V,2023arXiv230604837V}. 

The middle panel of Fig.~\ref{fig::gradients} shows the distribution of the metallicity gradients measured against the mean galactocentric distance, $R_g$. Interestingly, we see that the GSE stars show a weak but systematic negative metallicity gradient of $\displaystyle \approx -0.014^{-0.002}_{-0.022}$~dex/kpc, while in-situ and accreted stars, as we have seen before, are characterised by a normal distribution of gradients with the median values around zero. This is yet another example of Simpson's paradox in Galactic archaeology (see \citealt{2019MNRAS.487.3946M} for a list of cases related to the MW disk), in that a relation seen in a sub-sample (accreted) is lost when the total sample is considered. We note that the existence of a radial metallicity gradient in the GSE debris was reported previously by \cite{2022MNRAS.517.2787L}, who analysed the RR Lyrae stars in the MW halo and found the best value for the GSE merger debris gradient to be $-0.009$~dex/kpc, which is consistent with our estimate. We also note that slightly higher radial~($\sim -0.04$~dex/kpc) and vertical~($\sim -0.06$~dex/kpc) negative metallicity gradients of retrograde populations of the GSE debris have been detected by \cite{2020A&A...643A..69K}.

In a situation where we lack a large coverage of the halo tracers, using integrals of motion, particularly the total energy ($\rm E$), as a parameter to measure the metallicity gradient is indeed a more fundamental approach. As discussed in the introduction, several theoretical studies predict a systematic increase in the mean metallicity of stars with decreasing total energy. 

The right panel of Fig.~\ref{fig::gradients} shows the distribution of the radial metallicity gradients (\dfde) for all stars~(black), GSE stars~(blue), and in situ stars~(red). Similarly to the gradient measured along the mean galactocentric radius, we observe a negative metallicity gradient of approximately $\rm -0.28^{-0.3}_{-0.45}~dex/(10^5 km^2s^{-2})$ for the GSE stars, when using the total energy. \footnote{This is also a Simpson's paradox case, where this time, the relation (negative gradient) measured in both sub-populations (GSE and in-situ) deviate from the relation (no gradient) seen in the full sample.}
This negative gradient suggests that stars with lower total energy inside the MW halo tend to have higher metallicities. According to theoretical models~(see introduction), such a negative metallicity gradient may be a consequence of the gradual disruption of satellite galaxies with pre-existing radial metallicity gradients. In order to understand the origin and connection of the metallicity gradients we have discovered for the GSE merger debris, it is crucial to establish a link between these gradients and the parameters of its progenitor, such as its mass and initial metallicity distribution. In the next section, we provide the correspondence between the parameters of dwarf galaxies and their merger debris, in terms of the metallicity gradients.

\section{Modeling of $\nabla_E \FeH$}\label{sec::models}

\subsection{$N$-body merger simulations}
\subsubsection{$N$-body merger simulations: initial setup}
In order to study the redistribution of chemical abundances in the merger debris of an accreted galaxy we run 1000 $N$-body simulations of a single dwarf galaxy merger with a MW-like disc galaxy. In all simulations, the host galaxy is identical, and it was generated in a dynamical equilibrium using AGAMA with parameters from \cite{2017MNRAS.465...76M}, where the thin disk, thick disk, bulge and DM halo are represented by $10^5$, $10^5$, $10^4$ and $10^6$ particles, respectively. For each merger simulation, we generated a single component dwarf galaxy with a NFW density distribution, with the total masses $M_{dwarf}$ and a scale-length of $\rm (M_{dwarf}\times
10^{-11}/\Msun)^{0.6} \times 16$~kpc. This provides a reasonable range of masses and rotation curves for these systems~\citep{2021MNRAS.501.2279V}. In order to cover all possible parameter variations, we considered a flat distribution of masses in a log-scale from  $10^8$ to $10^{11.5}$~\Msun. Next, for each simulation, we randomly set up the initial position of a dwarf galaxy from $25$ to $100$~kpc from the host galaxy centre. Finally, the total velocity vector has the length of $0.5-0.9$ of the MW circular velocity at the location of the dwarf galaxy. 

We run the $N$-body models using our parallel version of the TREE-GRAPE code~\citep{2005PASJ...57.1009F} with multithread usage under the SSE and AVX instructions. In recent years we already used and extensively tested our hardware-accelerator-based gravity calculation routine in several galaxy dynamics studies where we obtained accurate results with a good performance~\citep{2018MNRAS.481.3534S, 2020A&A...638A.144K,2022A&A...663A..38K}. For the time integration, we used a leapfrog integrator with a fixed step size of $0.1$~Myr. In the simulation, we adopted the standard opening angle of $0.7$. All the simulations were run for $10$~Gyr.

We emphasise that our set of $N$-body models was not specifically designed to reproduce any particular features of the GSE debris. Instead, it provides valuable insights into the average relationship between the metallicity distribution inside dwarf galaxies and that observed in their merger debris. Hence, our approach aims to reveal general trends and characteristics of the GSE component based on statistical analyses of a diverse set of merger simulations. In contrast, \cite{2021ApJ...923...92N} pursued a different approach~\citep[see also][]{2020A&A...642L..18K}, running a large set of $N$-body mergers to find the best parameters for the GSE progenitor and its orbit to precisely reproduce the details of the accreted stellar halo of the MW. However, this approach comes with inherent problems. The 3D mass distribution of the MW, the amount of gas in both the MW and the GSE component and the orientation of the GSE galaxy relative to its orbit are not precisely constrained for the time of the merger~($8-11$~Gyr ago). While these uncertainties can introduce significant limitations and complexities when attempting to match specific details of the observed stellar halo, it's important to acknowledge that our models also do not consider the mass growth and gas content evolution of both the MW and the GSE galaxy. As a result, our conclusions may be subject to similar problems.

Nevertheless, to validate our results, which are based on the $N$-body models, in the subsequent section, we conduct a comparison with fully self-consistent HESTIA cosmological simulations. By doing so, we aim to enhance the robustness of our findings and address a broader range of physical processes and the cosmological context. This comparison will also help us better understand the GSE component and its connection to the dwarf galaxy progenitor.

\begin{figure*}
\begin{center}
\includegraphics[width=1.0\hsize]{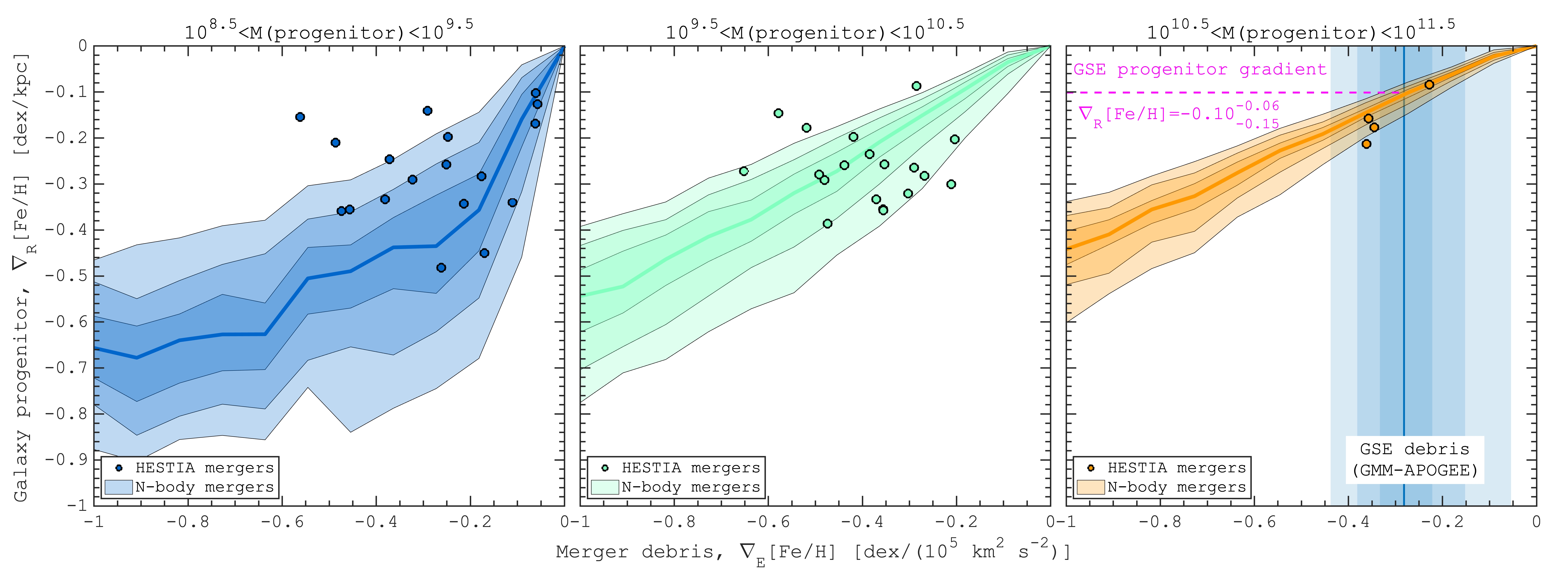}
\caption{Radial metallicity gradients of progenitor galaxies~(\dfdr) versus metallicity gradient measured as a function of the total energy of its debris~(\dfde). From left to right, different mass bins of the galaxy progenitor are shown as $10^{8.5-9.5}$~\Msun~(left), $10^{9.5-10.5}$~\Msun~(centre), $10^{10.5-11.5}$~\Msun~(right). Colour lines show the averaged trends based on the $N$-body merger simulation where different shades correspond to $(1-3)\sigma$ levels. The circles show the merger debris from the HESTIA simulations. In the right panel, the light blue area shows $(1-3)\sigma$ intervals for the GSE merger debris metallicity gradient~(see Section~\ref{fig::gradients}). The inferred radial metallicity gradient of the GSE progenitor galaxy and its median value, together with the lower and upper boundaries, are highlighted with the magenta line.}\label{fig::results}
\end{center}
\end{figure*}

\subsubsection{$N$-body merger simulations: metallicity gradients}
Here we analyse the results of our $N$-body merger simulations. In Fig.~\ref{fig::MMs_morphology} we present the stellar density distribution~(top) and the $\rm E-L_z$ space for 30 randomly selected simulations with the mass of the galaxy $\approx 10^9$~(first rows), $\approx 10^{10}$~(second rows) and $\approx 10^{11}$\Msun~(third rows). We see how much the total mass of the galaxy affects the merger debris structure and its integrals of motion space. For the low-mass bin, all the debris have a stream-like structure and are also confined in a small region of the $\rm E-L_z$ space~\citep{2000MNRAS.319..657H}. However, more massive galaxies result in smooth, but not fully featureless, merger debris which occupy a large area in the $\rm E-L_z$ coordinates~\citep{2017A&A...604A.106J,2023A&A...673A..86P}. We also note that massive mergers produce so-called energy wrinkles discovered in the \Gaia DR3 data~\citep{2023MNRAS.518.6200B} and attributed to multiple stripping episodes~\citep[but see also][for additional mechanisms]{2023arXiv230300008D,2023MNRAS.519..530D} clearly demonstrating that a single massive merger event can produce multiple clumps in the halo~\citep{2017A&A...604A.106J,2019MNRAS.487L..72G,2022arXiv220604522K}. 

Nevertheless, the primary objective of our simulations is to shed light on the relationship between the metallicity distribution inside dwarf galaxies and the structure of their merger debris. Since we utilise $N$-body simulations, we lack direct information about the metallicity of individual particles. However, this approach allows us to adopt reasonable metallicity gradients and observe how they are translated into energy coordinates after the merger.

To achieve this, for each $N$-body simulation, we assign a set of $30$ randomly chosen metallicity gradients in the range from $-1$ to $0$~dex/kpc. Consequently, for our set of $1000$ simulations, we have a total of $30,000$ models of metallicity gradient transformation. For each model, we measure the metallicity gradients as a function of galactocentric distance and as a function of total energy. The results are presented in Fig.~\ref{fig::MMs_grads}. In the left panel, we display the relations between $\dfde$ (merger debris) and $\dfdr$~(galaxy progenitor), while the right panel shows $\dfdr$~(merger debris) and $\dfdr$~(galaxy progenitor). The lines of different colours show the trends obtained by averaging the gradients in bins of the galaxy progenitor mass, as indicated by the colour bar. It is also important to emphasise that the trends include simulations with various orbital parameters of the merger, making use of statistical analysis to derive overall patterns and correlations. In the figure, we clearly observe a notable correlation, where a stronger metallicity gradient in the galaxy progenitor leads to a correspondingly stronger metallicity gradient for its merger debris. This finding indicates that the metallicity distribution within the dwarf galaxy still plays a significant role in shaping the resulting metallicity gradient of its merger debris after the interaction.

For low-mass mergers with masses below $10^{10}\Msun$, we observe a break in the relation. This break is likely due to the fact that these low-mass galaxies are not fully disrupted during the merger and are seen as coherent streams instead. As a consequence, these low-mass mergers do not follow the same trends as the more massive ones and represent a different category of interactions. It is important to note that this mass range is well below the predicted virial mass for the GSE progenitor~($10^{10}-10^{11}\Msun$, see introduction) and, thus, we are not aware of their behaviour. In Fig.~\ref{fig::MMs_grads} we see that the range of \dfde gradients is much larger compared to \dfdr; therefore, in order to avoid the degeneracy of our results in the following, we use \dfde gradients to constrain the parameters of the GSE progenitor. However, before using the relations in Fig.~\ref{fig::MMs_grads} to estimate the GSE progenitor metallicity gradient, we need to validate our $N$-body simulations results since they lack many physical processes relevant for the galaxy formation and evolution in the cosmological context.

\subsection{HESTIA simulations}
In this part of our work, we analyse the merger debris from three highest-resolution HESTIA simulations of the LG. Each simulation is tailored to reproduce a number of the LG properties~\citep{2020MNRAS.498.2968L}, including the massive disc galaxies resembling the MW and M31 analogues with the population of smaller satellites at $\rm z=0$. The HESTIA simulations are conducted using the AREPO code~\citep{2005MNRAS.364.1105S, 2016MNRAS.455.1134P} with the galaxy formation model developed by \cite{2017MNRAS.467..179G} and based on the Illustris model~\citep{2013MNRAS.436.3031V}. The initial conditions for the simulations are constrained using peculiar velocities from the Cosmicflows-2 catalogue ~ \citep{2013AJ....146...86T}, which were pre-processed with a bias-minimization technique based on the Wiener Filter/Constrained Realization algorithm~\citep{1991ApJ...380L...5H}. This technique is also combined with the Reverse Zeldovich Approximation~\citep{2013MNRAS.430..888D, 2013MNRAS.430..912D, 2013MNRAS.430..902D}.  For more details, we refer to the HESTIA simulations introductory paper~\citep{2020MNRAS.498.2968L}.

Highest resolutions HESTIA simulations are based on the re-runs of the low-resolution dark matter-only simulations where two overlapping $3.7$~Mpc ($\rm 2.5\, Mpch^{-1}$) spheres are drawn around the two main LG members at  $z = 0$ and then populated with $8192^3$ effective particles. The mass and spatial resolution achieved is $m_{dm} = 1.5\times 10^5$~\Msun, $m_{gas} = 2.2 \times 10^4$~\Msun\, and $\epsilon = 220$~pc. HESTIA simulations assume a cosmology consistent with the best fit values~\citep{2014A&A...571A..16P}: $\sigma_8 = 0.83$ and $\rm H0 = 100~h\, km\, s^{-1} Mpc^{-1}$ where $h = 0.677$. We adopt $\Omega_\Lambda = 0.682$ throughout and $\Omega_M = 0.270$ and $\Omega_b = 0.048$.  Thanks to the constrained initial conditions, HESTIA simulations provide a realistic framework for studying the effects of the LG environment on the dynamics and evolution of satellite galaxies~\citep{2022MNRAS.514.3612N, 2022MNRAS.516.4576D, 2023ApJ...946L..37N}, large-scale bulk motions of galaxy populations~\citep{2023MNRAS.523.5985P,2023MNRAS.523.2759S} and stellar/gas dynamics in and around the M31/MW analogues~\citep{2022arXiv221204515L, 2022MNRAS.517.6170B, 2022MNRAS.512.3717D}.

For our analysis, we focus on the most massive merger debris from six MW/M31 analogues, which have been extensively studied in a series of works by \cite{2022arXiv220605491K, 2022arXiv220604521K, 2022arXiv220604522K}. Specifically, we selected a total of $40$ mergers from these simulations, spanning a wide range of total masses from $10^{8.5}$ to $10^{11.5}$~\Msun. An important criterion for our selection was that the galaxy-progenitors of these mergers have negative metallicity gradients before the accretion event. This characteristic is crucial for our analysis as we aim to explore the influence of metallicity gradients in the progenitor galaxy on the resulting metallicity patterns in the merger debris.

\section{Inferring the GSE-progenitor metallicity gradient}\label{sec::results}

Fig.~\ref{fig::results} shows the combination of the measurements we conducted using our $N$-body merger simulations, HESTIA simulations and the GSE merger parameters we obtained using the GMM selection of its stars in the APOGEE. In particular, we show the relation between \dfde~(merger debris) and \dfdr~(progenitor galaxy) from the $N$-body simulations where filled areas of different shades correspond to $(1-3)\sigma$ around the median trend~(solid lines). The merger parameters from the HESTIA simulations are shown with the circles. Three panels correspond to three total mass ranges of the progenitor galaxy: $10^{8.5}-10^{9.5}$~\Msun~(left), $10^{9.5}-10^{10.5}$~\Msun~(middle), and $10^{10.5}-10^{11.5}$~\Msun~(right). 

First, we can see that the metallicity trends observed in our $N$-body simulations are consistent with those obtained from the HESTIA simulations for a given mass range. However, since the filling factor of HESTIA mergers in Fig.~\ref{fig::results} is rather low, we can use the relations from the $N$-body simulations to constrain the GSE progenitor metallicity gradient. To do so, we need to choose the total mass range of this galaxy. Previous works suggest that the virial mass of the GSE progenitor is $10^{10}-10^{11}~\Msun$~\citep{2018MNRAS.478..611B, 2020MNRAS.498.2472K,2021ApJ...923...92N, 2023arXiv230600770C}. These estimates are consistent with our $N$-body simulations which in Fig.~\ref{fig::MMs_morphology} show the morphology of the debris, suggesting that systems with lower masses should be seen as streams and only massive galaxies accretion result in a smooth merger debris structure covering large volumes of the $\rm E-L_z$ space, as we see in the analysis of the APOGEE data~(see Fig.~\ref{fig::validation}). Therefore, our choice is the right panel in Fig.~\ref{fig::results}, where the grey shaded area reflects the distribution of the \dfde gradients obtained in the GMMs with the mean value of $\rm -0.1 dex/(10^5 km^2 s^{-2})$~(see Fig.~\ref{fig::gradients}). From these data points, we obtain the GSE progenitor radial metallicity gradient of $-0.1_{-0.15}^{-0.06}$~dex/kpc. Interestingly, this value is just a little lower compared to the MW metallicity gradient of $\approx -0.13$~dex/kpc~\citep{minchev18, 2022arXiv221204515L, 2023MNRAS.tmp.1561R} at the time of the GSE accretion.


This estimation of the radial metallicity gradient of the GSE progenitor represents a novel and crucial piece of information in understanding its evolution and its relation to known populations of galaxies. This new finding provides valuable insights into the chemical enrichment processes and the formation history of the GSE progenitor. Moreover, it opens up new avenues for research and advances our understanding of the complex processes that govern the formation and evolution of galaxies in the universe.

\section{Discussion. GSE progenitor galaxy: not a giant but not a dwarf}\label{sec::discussion}

In our analysis, we demonstrated that the radial metallicity gradient of the GSE progenitor is approximately $-0.1$~dex/kpc. While this value is not extreme among the population of nearby dwarf galaxies~\citep[see, e.g. a recent compilation in][showing the range from $\approx -0.5$ to $\approx 0$~dex/kpc]{2022A&A...665A..92T}, what stands out is the stellar mass of the GSE progenitor, which falls within the range of $(0.03-0.7)\times 10^{10}$ \Msun, in comparison to the satellite galaxies in the LG~($10^4-10^8$~\Msun, see, e.g. \cite{2015ApJ...804..136W,2022NatAs...6..659B}). The GSE progenitor's stellar mass is notably higher than most LG dwarf galaxies. However, there are LG dwarf galaxies with comparable stellar masses, such as Sgr and LMC. The distinguishing factor is that these objects are  galaxies that have been actively forming stars over the last several billion years, which is not the case for the GSE progenitor. This discrepancy in star formation history sets the GSE progenitor apart from other dwarf galaxies in the LG and emphasizes its quite unique nature, which is also evident from its chemical abundance composition~\citep{2020MNRAS.497.1236M,2021ApJ...923..172H}.

The massive GSE progenitor seems not to be a typical dwarf galaxy despite its mass being rather low compared to the present-day MW. However, this is not a fair comparison because the GSE progenitor ended its 'individual' evolution at redshift $z\approx2$. At that time, the stellar mass of the MW was about  $(0.5-2)\times 10^{10}$~\Msun~\citep[see e.g.,][ and references in Introduction]{2015A&A...578A..87S,2018A&A...618A..78H}. This suggests a stellar mass ratio between the GSE progenitor and the young MW from $1:4$ to $1:100$. These numbers can be constrained further if one considers that masses of galaxies correlate with the number of their globular clusters at $z=0$~\citep{2018MNRAS.481.5592F, 2017ApJ...836...67H, 2005ApJ...623..650K, 2020AJ....159...56B}, assuming that their disruption rate weakly depends on their origin. The association of the GCs with their progenitor galaxies is not unambiguous~\citep{2023A&A...673A..86P} and different methods provide the following ratio between GSE and MW: $\approx 1:2$~\citep{2019A&A...630L...4M}, $\approx 1:5$~\citep{2020MNRAS.498.2472K}, $\approx 1:3$~\citep{2022ApJ...926..107M}. Therefore, indeed the mass of the GSE-progenitor is lower but comparable to the MW at $z\approx2$, but we do not call proto-MW a dwarf galaxy. 

Besides the population of GCs, there is another crucial system closely tied to massive MW-type galaxies, which is the population of their satellites orbiting around their hosts. Cosmological simulations have provided significant insights into the hierarchical nature of galaxy formation and evolution, demonstrating that even before redshift $z=2$, more massive systems are typically surrounded by a large number of smaller satellite systems. In this context, it is quite reasonable to assume that both the MW and the GSE progenitor had their own populations of satellite galaxies. During the MW-GSE merger, satellites of the GSE would have been accreted as well, contributing to the overall population of the MW satellites. These accreted satellites, previously satellites of the GSE progenitor, should retain a memory of their initial orbit and be seen as kinematically distinct satellite subsystems in the MW. The only coherent structures around galaxies are the planes of satellites. Although it is quite natural to assume that the planes of satellites can be the results of mergers with similar systems a long time ago, there is not yet evidence in favour of such a connection in the MW. Nevertheless, it would be interesting to link such phenomena with the GSE merger in the MW, also because a few known planes of satellites are found in galaxies that have themselves experienced recent, massive mergers~(M31, Centaurus A). Therefore, if the GSE merger in the MW is shown to be connected to the planes of satellite galaxies, it could provide a brilliant manifestation of the hierarchical galaxy formation in the CDM model.

\section{Summary}\label{sec::summary}

In this study, we adopted a comprehensive approach by combining the analysis of observational data related to the MW stellar halo with simulations of galaxy mergers. This unique combination allowed us to draw inferences and insights into the parameters of the progenitor galaxy responsible for the GSE merger event. We summarize the main results as follows.

\begin{itemize}
    \item We employed unsupervised Gaussian Mixture Models to classify chemically pre-selected stars in the $\FeH<-0.5$ dex and $\AlFe<-0.05$ dex region. This selection targeted stars as potential candidates for the GSE merger debris. The GMM-based classification provided us with a probability score for each star, indicating the likelihood of it belonging to the GSE component. The analysis of the likelihood distribution in chemical abundance and kinematic $\rm E-L_z$ coordinates reveals a substantial contribution of in situ stars. These stars, with no bulk rotation, follow the trends expected for the low-metallicity tail of the high-$\alpha$ sequence and correspond to proto-MW disc populations.

    \item We found that the GSE stars do not exhibit a significant metallicity gradient when considered as a function of their instantaneous galactocentric distances. However, a notable finding emerged when we investigated the relationship between the mean galactocentric distance ($R_g$) and the metallicity of the GSE debris. We discovered a shallow yet statistically significant radial metallicity gradient of $\nabla_{R_g} = -0.014_{-0.02}^{-0.002}$~dex/kpc when considering the mean galactocentric distance as the parameter. This result suggests that, on average, the metallicity of the GSE merger debris exhibits a subtle decrease as one moves to larger galactocentric distances. We argued however that the metallicity gradient of the GSE debris over the total energy is a more valuable piece of information in our study. We estimated this gradient to be $\rm \dfde=-0.28_{-0.45}^{-0.03}~dex/(10^5 km^2 s^{-2})$, which allowed us to gain a better understanding of how the chemical composition varies with the properties of its galaxy progenitor.

    \item We conducted a series of 1000 $N$-body simulations of galaxy mergers, allowing us to establish a relationship between the radial metallicity gradient inside the galaxy and the metallicity gradients of its merger debris. This relationship was further validated through self-consistent HESTIA simulations of the LG in a cosmological context. Using this information, we constrained the radial metallicity gradient of the GSE galaxy progenitor to be approximately $\approx -0.1^{-0.06}_{-0.15}$ dex/kpc. This value closely resembles the metallicity gradients observed in nearby dwarf galaxies with old stellar populations. However, it is important to note that despite the similarity in the metallicity gradients, the GSE progenitor was much more massive compared to the present-day local dwarfs. This disparity in mass highlights the distinct nature of the GSE progenitor compared to nearby dwarf galaxies with similar metallicity gradients. Therefore, the GSE progenitor's massive nature has significant implications for its formation and evolutionary history, suggesting a different assembly process compared to typical nearby dwarf galaxies. 

\end{itemize}

Galaxy merger analysis is a powerful approach in the context of Galactic archaeology that combines observational data, theoretical models, and simulations to unravel the history and evolution of galaxies. Studying the remnants of galactic mergers, such as the GSE merger in the MW, provides unique insights into the formation and assembly of stellar halos, their progenitor galaxies and their dynamical interactions with the MW. Coming spectroscopic data from large-scale surveys, like 4MOST~\citep{2019Msngr.175....3D}, MOONS~\citep{2020Msngr.180...18G}, WEAVE~\citep{2023MNRAS.tmp..715J}, will soon increase the amount, quality and dimensionality of data about MW stellar populations and thus serve as a crucial tool to further understand the accretion history, mass growth, and chemical enrichment processes that shaped the MW over cosmic time, offering valuable contributions to our broader understanding of galaxy formation and evolution.

\begin{acknowledgements}
This research made use of data from the European Space Agency mission Gaia
(\url{http://www.cosmos.esa.int/gaia}), processed by the Gaia Data
Processing and Analysis Consortium (DPAC,
\url{http://www.cosmos.esa.int/web/gaia/dpac/consortium}). Funding for the
DPAC has been provided by national institutions, in particular the
institutions participating in the Gaia Multilateral Agreement.
I.M. acknowledge support by the Deutsche Forschungsgemeinschaft under the grant MI 2009/2-1.

Funding for the Sloan Digital Sky Survey IV has been provided by the Alfred P. Sloan Foundation, the U.S. Department of Energy Office of Science, and the Participating Institutions. SDSS-IV acknowledges support and resources from the Center for High Performance Computing at the University of Utah. The SDSS website is www.sdss.org. SDSS-IV is managed by the Astrophysical Research Consortium for the Participating Institutions of the SDSS Collaboration including the Brazilian Participation Group, the Carnegie Institution for Science, Carnegie Mellon University, Center for Astrophysics | Harvard \& Smithsonian, the Chilean Participation Group, the French Participation Group, Instituto de Astrofísica de Canarias, The Johns Hopkins University, Kavli Institute for the Physics and Mathematics of the Universe (IPMU)/University of Tokyo, the Korean Participation Group, Lawrence Berkeley National Laboratory, Leibniz Institut für Astrophysik Potsdam (AIP), Max-Planck-Institut für Astronomie~(MPIA Heidelberg), Max-Planck-Institut für Astrophysik (MPA Garching), Max-Planck Institut für Extraterrestrische Physik (MPE), National Astronomical Observatories of China, New Mexico State University, New York University, University of Notre Dame, Observatário Nacional/MCTI, The Ohio State University, Pennsylvania State University, Shanghai Astronomical Observatory, United Kingdom Participation Group, Universidad Nacional Autónoma de México, University of Arizona, University of Colorado Boulder, University of Oxford, University of Portsmouth, University of Utah, University of Virginia, University of Washington, University of Wisconsin, Vanderbilt University, and Yale University.

\end{acknowledgements}

\bibliographystyle{aa}
\bibliography{refs}

\end{document}